\documentclass[article, aps, 12pt, floatfix, longbibliography, a4paper]{revtex4-2}

\usepackage{graphicx}
\usepackage{epstopdf}

\usepackage{color}
\usepackage{amsmath,bm}
\usepackage[utf8]{inputenc}
\usepackage{xcolor}
\usepackage{physics}
\usepackage{mathcomp}
\usepackage{comment}
\usepackage{ulem}
\usepackage[resetlabels,labeled]{multibib}

\begin{document}

\title[]{\textcolor{black}{Emergent electric field induced by dissipative sliding dynamics\\ of domain walls in a Weyl magnet}}


\author{Rinsuke Yamada$^{1}$}\email{ryamada@ap.t.u-tokyo.ac.jp}
\author{Daichi Kurebayashi$^{2,3}$}
\author{Yukako Fujishiro$^{3,4}$}
\author{Shun Okumura$^{1}$}
\author{Daisuke Nakamura$^{3}$}
\author{Fehmi S. Yasin$^{3,5}$}
\author{Taro Nakajima$^{\textcolor{black}{3,6,7}}$}
\author{Tomoyuki Yokouchi$^{3}$}
\author{Akiko Kikkawa$^{3}$}
\author{Yasujiro Taguchi$^{3}$}
\author{Yoshinori Tokura$^{1,3,8}$}
\author{Oleg A. Tretiakov$^{2}$}
\author{Max Hirschberger$^{1,3}$}\email{hirschberger@ap.t.u-tokyo.ac.jp}

\affiliation{$^{1}$Department of Applied Physics and Quantum-Phase Electronics Center, The University of Tokyo, Bunkyo, Tokyo 113-8656, Japan}
\affiliation{$^{2}$School of Physics, The University of New South Wales, Sydney, New South Wales 2052, Australia}
\affiliation{$^{3}$RIKEN Center for Emergent Matter Science (CEMS), Wako, Saitama 351-0198, Japan}
\affiliation{$^{4}$RIKEN Cluster for Pioneering Research, Wako, Saitama 351-0198, Japan}
\affiliation{$^{5}$Center for Nanophase Materials Sciences, Oak Ridge National Laboratory, Oak Ridge, Tennessee 37831, United States}
\affiliation{$^{6}$The Institute for Solid State Physics, The University of Tokyo, Kashiwa 277-8581, Japan}
\affiliation{\textcolor{black}{$^{7}$Institute of Materials Structure Science, High Energy Accelerator Research Organization, Tsukuba, Ibaraki 305-0801, Japan}}
\affiliation{$^{8}$Tokyo College, The University of Tokyo, Tokyo 113-8656, Japan}

\maketitle

\textbf{The dynamic motion of topological defects in magnets induces an emergent electric field, as exemplified by the continuous flow of skyrmion vortices. However, the electrodynamics underlying this emergent field remains poorly understood. In this context, magnetic domain walls —one-dimensional topological defects with two collective modes, sliding and spin-tilt—offer a promising platform for exploration. Here, we demonstrate that the dissipative motion of domain walls under oscillatory current excitation generates an emergent electric field. We image domain patterns and quantify domain wall length under applied magnetic fields in mesoscopic devices based on the magnetic Weyl semimetal NdAlSi. These devices exhibit exceptionally strong domain wall scattering and a pronounced emergent electric field, observed in the imaginary component of the complex impedance. Spin dynamics simulations reveal that domain wall sliding dominates over spin tilting, where the phase delay of the domain wall motion with respect to the driving force impacts the emergent electric field. Our findings establish domain-wall dynamics as a platform for studying emergent electromagnetic fields and motivate further investigations on the coupled motion of magnetic solitons and conduction electrons.
}\\

Notice: This manuscript has been authored by UT-Battelle, LLC, under contract DE-AC05-00OR22725 with the US Department of Energy (DOE). The US government retains and the publisher, by accepting the article for publication, acknowledges that the US government retains a nonexclusive, paid-up, irrevocable, worldwide license to publish or reproduce the published form of this manuscript, or allow others to do so, for US government purposes. DOE will provide public access to these results of federally sponsored research in accordance with the DOE Public Access Plan (https://www.energy.gov/doe-public-access-plan).

\newpage

\begin{center}
\Large{Main text}
\end{center}

Emergent electromagnetism in solids provides a unified description for the interplay of moving charges and quantum-mechanical spin degrees of freedom. In the limit where the magnetization is seen as a continuous field in space, dynamic motion of magnetic textures causes an emergent electric field that represents an infinitesimal solid angle swept by spins in space and time,
\begin{equation}
\label{eq:emergent_electric_field}
e_i = \frac{h}{2 \pi e} \bm{n} \cdot (\partial_i \bm{n} \times \partial_t \bm{n}),
\end{equation}
where $h$ and $e$ are Planck's constant and the electron charge, and $\bm{n}$ is a unit vector parallel to the local spin direction~\cite{GEVolovik1987, GTatara2004, SEBarnes2007, JKishine2010, NNagaosa2019}. This emergent electric field (EEF) is a potentially large cousin of Maxwell's electromagnetic fields, created by collective many-body behaviour in solids~\cite{ANeubauer2009, TSchulz2012, NNagaosa2012}.
Whereas observing the EEF under alternating-current (a.c.) drive was originally proposed as a probe of collective dynamics in helimagnets~\cite{NNagaosa2019, TYokouchi2020}, the EEF of localised topological defects -- magnetic domains walls (DWs), spin vortices, and stable solitons such as skyrmions -- is less explored but perhaps more technologically relevant~\cite{SAYang2009, YShoka2023, YMatsushima2024, ZZhang2024, SParkin2008, Finocchio2016, AFert2017}. 

Dynamics of topological defects at high current densities has been extensively probed by imaging and electron transport~\cite{GCatalan2012, Shibata2011}, and the EEF of depinned and uniformly flowing magnetic solitons has been observed~\cite{TSchulz2012,Birch2024, KLitzius2007}. Much less is known about a.c.~dynamics \textcolor{black}{of magnetic solitons in the pinned regime} and the \textcolor{black}{resulting} EEF in the pinned regime. When an oscillatory current drives the soliton, \textcolor{black}{energy is transferred to the spin system}~\cite{SFuruta2023}; a part of this energy input is lost or dissipated by friction, but some energy is stored in kinetic and potential energy of the spin system and can be later returned to the electron gas via spin-dynamical creation of the EEF. In experiments, the complex impedance $\Im \rho_{xx}$ serves as a sensitive yardstick for the EEF's strength.
Detection of the EEF by complex impedance measurements with an a.c.~excitation current is \textcolor{black}{thus} widely applicable to various \textcolor{black}{types of} magnetic solitons in the pinned regime\textcolor{black}{, but the power of the technique was not realized until recently~\cite{NNagaosa2019} and the EEF has not yet been used as a probe of friction, or energy dissipation, in soliton dynamics}.

\begin{figure}[h]%
\centering
\includegraphics[trim=0cm 0.1cm 0.0cm 0.0cm,clip,width=0.95\textwidth]{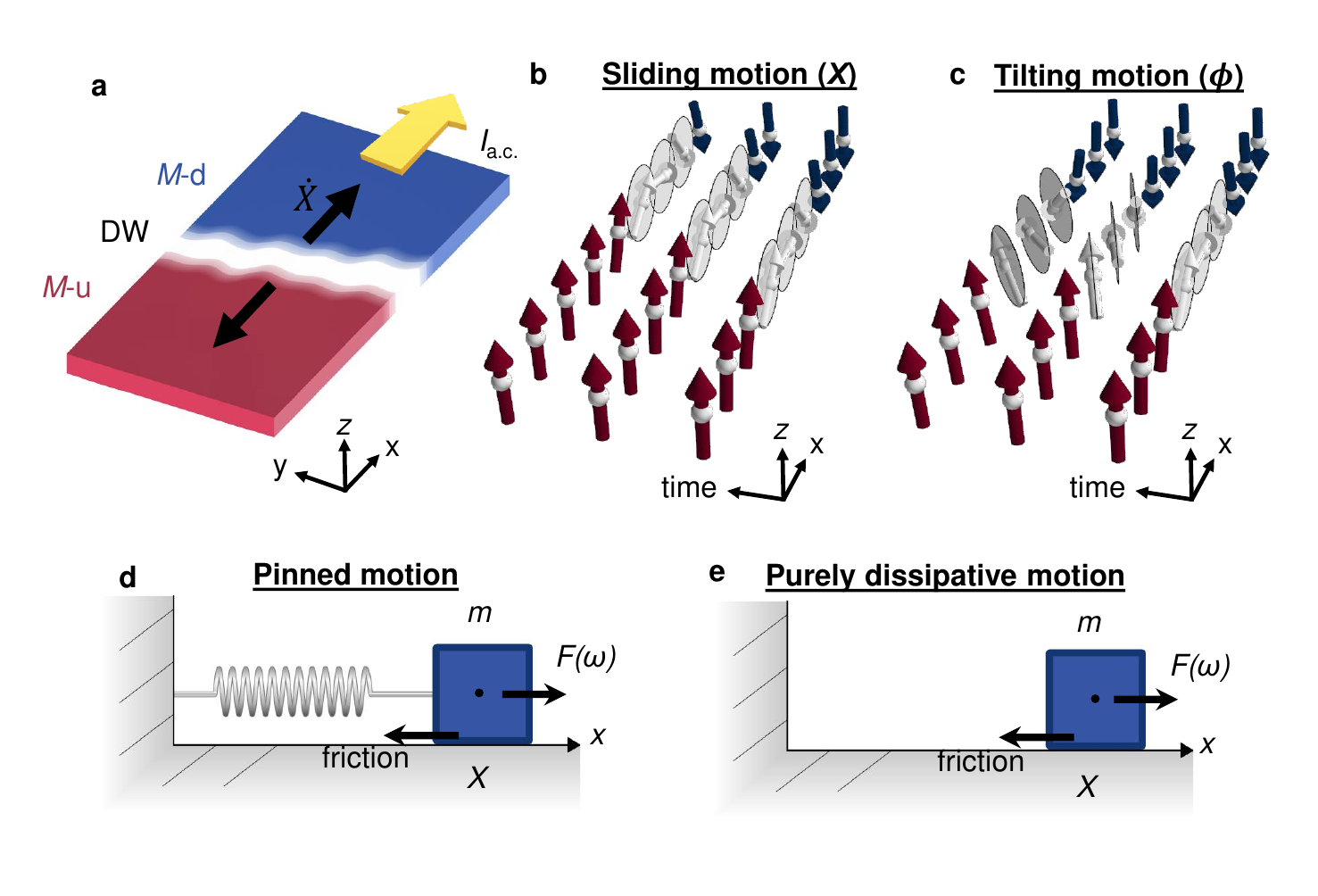}
\caption{\textbf{Schematics for a magnetic domain wall (DW) and its dissipative motion.} \textbf{a}, DW motion excited by an a.c.~electric current $I_\mathrm{a.c.}$ (yellow arrow). Red, blue, and white regions indicate domains with positive and negative net magnetization, as well as the DW region. Black arrows indicate the sliding motion of a magnetic DW via the time derivative of the sliding mode $\dot{X}$.
\textbf{b},\textbf{c}, Schematics of the time evolution of the sliding motion ($X$) and spin-tilting motion ($\phi$) under $I_\mathrm{a.c.}$ along the $x$ axis. We illustrate the sliding motion in the case of $\phi = 0$ in panel \textbf{b}, while $X$ is set to zero in panel \textbf{c}. 
\textbf{d}, Mass on a spring as a driven harmonic oscillator, subject to friction and an external force $F$ induced by excitation current $I_{a.c.}(\omega)$. \textbf{e}, Dissipative oscillator without reactive spring-force. 
}\label{Fig1}
\end{figure}

Focusing on DWs as a simple solitonic structure, Fig.~\ref{Fig1}\textbf{a}-\textbf{c} introduces two essential collective coordinates to model the EEF and $\Im\rho_{xx}$~\cite{NNagaosa2019, DKurebayashi2019, DKurebayashi2021,YYamane2022, YAraki2023, TOh2024}: the sliding mode $X$ and the spin-tilting mode $\phi$. 
Here, $X$ is the position of the centre of a magnetic DW, while $\phi$ is the tilt angle of the DW's centre.
In terms of these, analytic expressions for the EEF have been obtained as a linear response to the applied electric current $I_\mathrm{a.c.}$~\cite{NNagaosa2019, DKurebayashi2021, YAraki2023, TOh2024}.
Key challenges in this field are then not only observing a complex impedance for magnetic solitons, ideally supported by real-space imaging experiments, but also revealing the energetic balance in the soliton's trapped dynamics between energy of spin-tilting $\phi$, energy of the sliding mode $X$, and friction 
(Fig.~\ref{Fig1}\textbf{d},\textbf{e}). 
A thorough theoretical description also requires extending existing models into the current-nonlinear regime. 

With the explicit goal of realizing the simple limit of dominant sliding motion of DWs caused by strong dissipation, we choose the Weyl magnet NdAlSi, where N\'{e}el and Bloch DWs 
are nearly degenerate \textcolor{black}{and the spin rotation plane at the DW can easily tilt}~\cite{JGaudet2021, BJuba2024}. We find that in this limit of 'easy' spin tilting, the dynamics of the DW can be described by an effective model for the sliding motion $X$ only, where the dominant role of friction in the pinned, oscillatory motion of DWs can be detected by the negative sign of $\Im\rho_{xx}$. We argue that $\Im\rho_{xx}$ measurements, combined with modeling the EEF based on the Landau-Lifshitz-Gilbert equation, are a new way to quantify local dissipation in the process of soliton motion in magnets.
\\

\textbf{Large domain wall resistivity and imaginary impedance}

\begin{figure}[h]%
\centering
\includegraphics[trim=0.0cm 0.0cm 0.0cm 0.0cm,clip,width=0.95\textwidth]{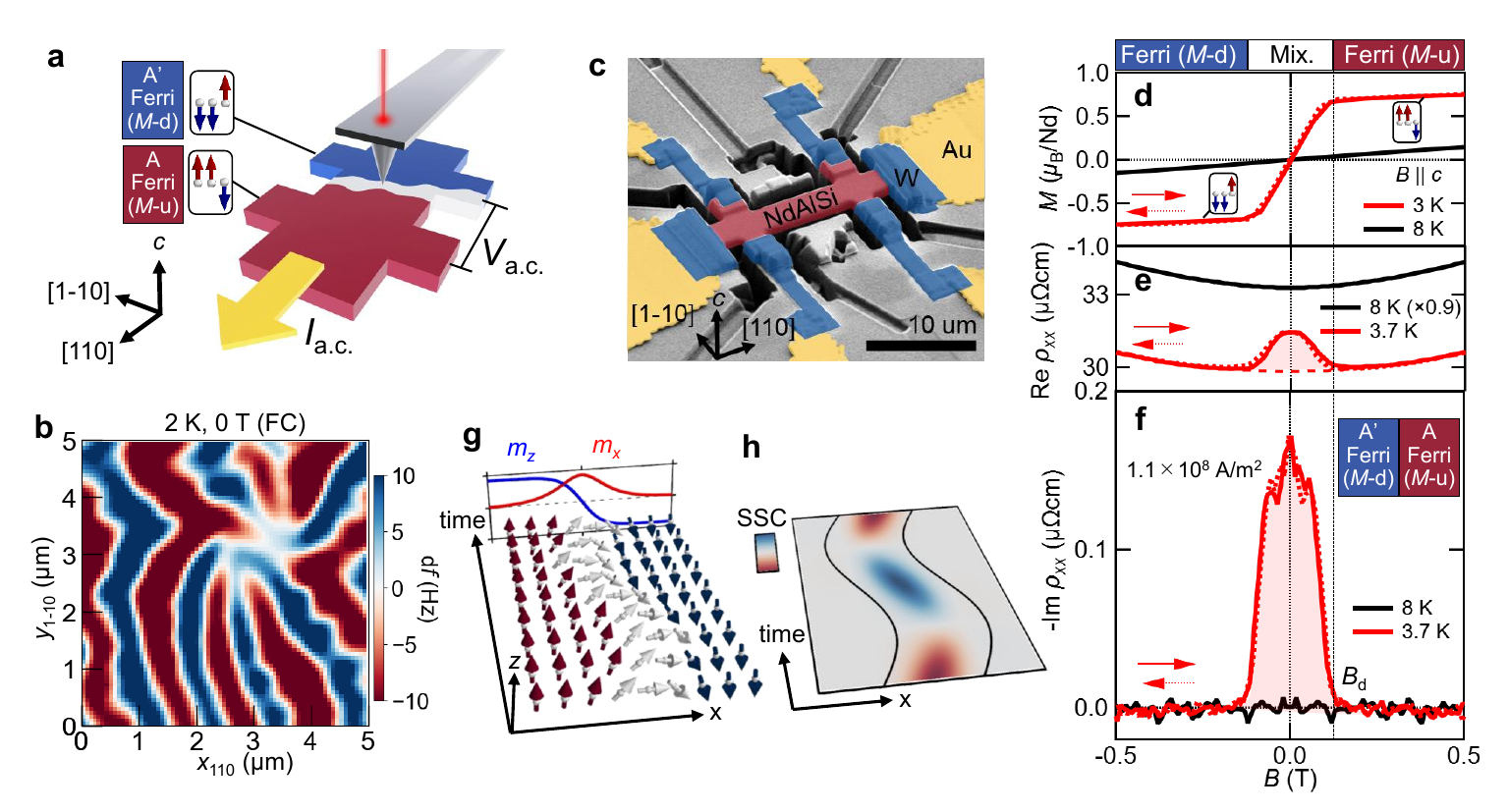}
\caption{\textbf{Domain wall (DW) resistance and emergent electric field (EEF) in Weyl magnet NdAlSi.} $\mathbf{a,}$ Illustration of magnetic force microscopy (MFM) measurements. The magnetic stray field from the sample is detected optically as a variation of the resonance frequency of the cantilever (Methods). $\mathbf{b,}$ MFM image of the mesoscopic device of NdAlSi at zero external field and $T=2\,\mathrm{K}$ under field-cooling conditions (Methods). In panels $\mathbf{a,b}$, the area with negative (positive) net magnetization is colored in dark blue (dark red), respectively, while the white area is the magnetic DW. $\mathbf{c,}$ False-colour image of a micrometer-sized device of NdAlSi fabricated by the focused ion beam (FIB) technique for the transport measurements (Methods). $\mathbf{d,}$ Magnetization ($M$) curves measured at $3\,\mathrm{K}$ ($<T_\mathrm{C}$) and $8\,\mathrm{K}$ ($>T_\mathrm{C}$). $\mathbf{e,}$ Magnetoresistivity ($\mathrm{Re}\rho_{xx}$) of the FIB device of NdAlSi. The dashed line is a fit to the $B^2$ term from the Lorentz force, while the area highlighted in red corresponds to DW-scattering ($\mathrm{Re} \, \Delta \rho_{xx}^\mathrm{DW}$). $\mathbf{f,}$ Imaginary impedance $\Im \rho_{xx}$ originating from DWs. The vertical dotted line in panels \textbf{d-f} indicates the magnetic field where the DWs disappear ($B_\mathrm{d}$). \textcolor{black}{$\Im \rho_{xx}$ is enhanced below $B_\mathrm{d}$, as highlighted in red in panel \textbf{f}.} $\mathbf{g,h,}$ Schematic time-evolution of the spin moments at a magnetic domain wall under excitation currents and the calculated scalar spin chirality (SSC) in the space-time domain (see Supplementary Fig. 4). Finite, time-alternating SSC results in an EEF detected by the imaginary impedance.
}\label{Fig2}
\end{figure}

The $R$AlSi/$R$AlGe ($R$ = rare earth) family crystallises in the tetragonal $I4_{1}md$ space group without inversion centre, leading to a Weyl semimetal state~\cite{SYXu2017,JGaudet2021,RYamada2024,BJuba2024} with high carrier mobility and a sizable spin-charge coupling (see Extended Data Fig. 1, Supplementary Table 1, and Supplementary Note 3). \textcolor{black}{We investigate the domain wall (DW) response to an alternating excitation current $I_\mathrm{a.c.}$ in} NdAlSi, which shows ferrimagnetic order of Ising character below $T_\mathrm{C} \sim 7$ K~\cite{JGaudet2021}. Domains of predominantly up ($\uparrow \uparrow\downarrow$) and down ($\downarrow\downarrow\uparrow$) magnetization are almost equally populated at zero magnetic field (Fig.~\ref{Fig2}\textbf{a}, Extended Data Fig. 2). 
The domain patterns are investigated by magnetic force microscopy (MFM, Methods, Supplementary Fig. 1). As shown in Fig.~\ref{Fig2}$\mathbf{a}$, the MFM cantilever is a mechanical resonator whose characteristic frequency shifts by an amount $df(x,y)$ due to the change in the stray field emanating from the sample, when the scanning position of the cantilever is $(x,y)$. Figure \ref{Fig2}\textbf{b} shows the observed domain pattern in zero magnetic field (Methods for field-cooling procedure). A zig-zag DW (white area) separates two stripe-domains with positive and negative net magnetization, which are coloured in dark red and dark blue, respectively. 

For detection of the EEF from DWs, we fabricate thin-plate devices using the focused-ion beam (FIB) technique (Fig.~\ref{Fig2}\textbf{c}, Methods). NdAlSi is suitable for this study, as we can obtain high-quality devices without much material deterioration during fabrication. In the following, we refer to the real and imaginary parts of the complex resistivity $\rho_{xx} = V_\mathrm{a.c.}/I_\mathrm{a.c.} \times S/l$ ($S$: cross-sectional area, $l$: voltage terminal distance) as "resistivity" and "imaginary impedance", which are in-phase and out-of-phase to the \textcolor{black}{oscillatory} excitation $I_\mathrm{a.c.}$, respectively. In Fig.~\ref{Fig2}\textbf{e}, the resistivity $\Re\rho_{xx}$ monotonically increases as the square of the magnetic field $B^2$ above $T_\mathrm{C}$, in consequence of the classical Lorentz force on conduction electrons. However, during the magnetization process (see Fig.~\ref{Fig2}\textbf{d}), sizable DW scattering appears in $\Re\rho_{xx}$ below the magnetic field $B_\mathrm{d}\sim0.13\,\mathrm{T}$, indicated by red shading in Fig.~\ref{Fig2}\textbf{e}. Above $B_\mathrm{d}$, the DWs disappear as the domains with positive magnetization take over the entire sample volume. We extract the DW contribution $\Re \rho_{xx}^\mathrm{DW}$ from $\Re \rho_{xx}$ by subtracting the $B^2$ term as depicted in Fig.~\ref{Fig2}\textbf{e}. This $\Re \rho_{xx}^\mathrm{DW}$ reaches $1.6\,\mathrm{\mu \Omega cm}$ or about $5\,\%$ of the total resistivity, which is one order of magnitude larger than $\Re \rho_{xx}^\mathrm{DW}$ in conventional ferromagnets (see Supplementary Note 11). Indeed, the DW resistivity in NdAlSi can be further enhanced up to $10\,\%$ by utilising the zero field cooling (ZFC) procedure (Supplementary Table 3). The experimental findings are consistent with theories of Weyl semimetals with enhanced DW scattering, caused by their small Fermi surface cross-section and two aspects of relativistic spin-orbit coupling: Namely, spin-momentum locking of conduction electrons and the sizable shift of the Fermi surface in momentum space \textcolor{black}{between the two magnetic domains with opposite net magnetization}, as caused by the Zeeman effect~\cite{KKobayashi2018, YOminato2017, Nguyen2006, XMei2021, YAraki2020, DXLi2020}.

Encouraged by the clear visibility of DWs in electron transport, we examine the imaginary impedance $\Im \rho_{xx}$ in Fig.~\ref{Fig2}\textbf{f}. A bell-shaped anomaly below $B_\mathrm{d}$, as well as the absence of the signal above $T_\mathrm{C}$, suggests that a large EEF signal is produced by DWs between the two types of domains visualised in Fig.~\ref{Fig2}\textbf{b}. In our theoretical model, $\Im \rho_{xx}$ detects the phase-delayed EEF response created by current-induced dynamics of the collective coordinates $X$ and $\phi$ of a DW, defined in Fig.~\ref{Fig1}\textbf{b},\textbf{c}. 
When the excitation current is moderately small and DWs remain pinned to local defects, the coupled equations of motion for $X$ and $\phi$, Eqs.~(\ref{EqofMotion_X},\ref{EqofMotion_phi}) of Methods, describe the oscillatory motion of the DW driven by spin-transfer torque. Focusing on the equations of motion for the sliding mode $X$, we obtain
\begin{equation}
\label{EqofMotion_X_twodots}
\ddot{X} + \alpha (\Omega_\mathrm{int} + \Omega_\mathrm{ext}) \dot{X} + (1 + \alpha^2) \Omega_\mathrm{int} \Omega_\mathrm{ext} X = \beta \Omega_\mathrm{int} \nu + \frac{1+\alpha\beta}{1 + \alpha^2} \dot{\nu},
\end{equation}
which is composed of the mass, friction, and potential terms on the left and the driving force from the excitation current on the right. Here, $\Omega_\mathrm{int}$, $\Omega_\mathrm{ext}$, $\alpha$ are the intrinsic pinning frequency originating from the magnetic Dzyaloshinskii-Moriya interaction, extrinsic pinning frequency induced by the attraction of a DW to magnetic defects, and the Gilbert damping parameter quantifying energy dissipation in spin dynamics, respectively. The dimensionless constants $\beta$ and $\nu$ quantify non-adiabatic spin dynamics and the applied electric current, respectively (Methods). From this current-driven dynamics \textcolor{black}{of $X$ and $\phi$}, we calculate the EEF $e_i \sim \bm{n} \cdot (\partial_i \bm{n} \times \partial_t \bm{n})$ of a DW and $\Im\rho_{xx}$ as a spin-twisting in space-time, see Fig.~\ref{Fig2}\textbf{g,h}. 
We exclude contributions to $\mathrm{Im}\,\rho_{xx}$ from temperature oscillations in Supplementary Fig. 3 and Supplementary Note 9~\cite{JIeda2021, YYamane2022}.
\\

\textbf{Figure of merit and negative sign of impedance}\\
At this point, it is convenient to introduce the quality factors $Q = \Im \rho_{xx} / \Re \rho_{xx}$ and $Q^\mathrm{DW} = \Im \rho_{xx} / \Re \rho_{xx}^\mathrm{DW}$, which describe the relative magnitude of the EEF signal. According to Fig.~\ref{Fig2}\textbf{f}, $\Im \rho_{xx}$ reaches $0.15\, \mathrm{\mu\Omega cm}$ when the DW resistivity is $\Re \rho_{xx}^\mathrm{DW} = 1.0\, \mathrm{\mu \Omega cm}$, while the total resistivity is $31\, \mathrm{\mu\Omega cm}$. Therefore, the quality factor $Q = 0.15 / 31 \sim 0.5 \,\%$ of the DW EEF in NdAlSi is comparable to the values observed in bulk helimagnets~\cite{TYokouchi2020,NNagaosa2019}, even though the present EEF of DWs emanates from a limited volume fraction of the device. Meanwhile, $Q^\mathrm{DW}=0.15 / 1.0 \sim 15\,\%$ represents a measure of a more perfect crystal, without scattering from quenched (chemical) disorder. We can see that the EEF from spin dynamics of DWs in NdAlSi is comparable even to the enhanced DW scattering $\Re \rho_{xx}$ in this magnetic Weyl semimetal.

\begin{figure}[h]%
\centering
\includegraphics[trim=0.0cm 0.0cm 0.0cm 0.0cm,clip,width=0.95\textwidth]{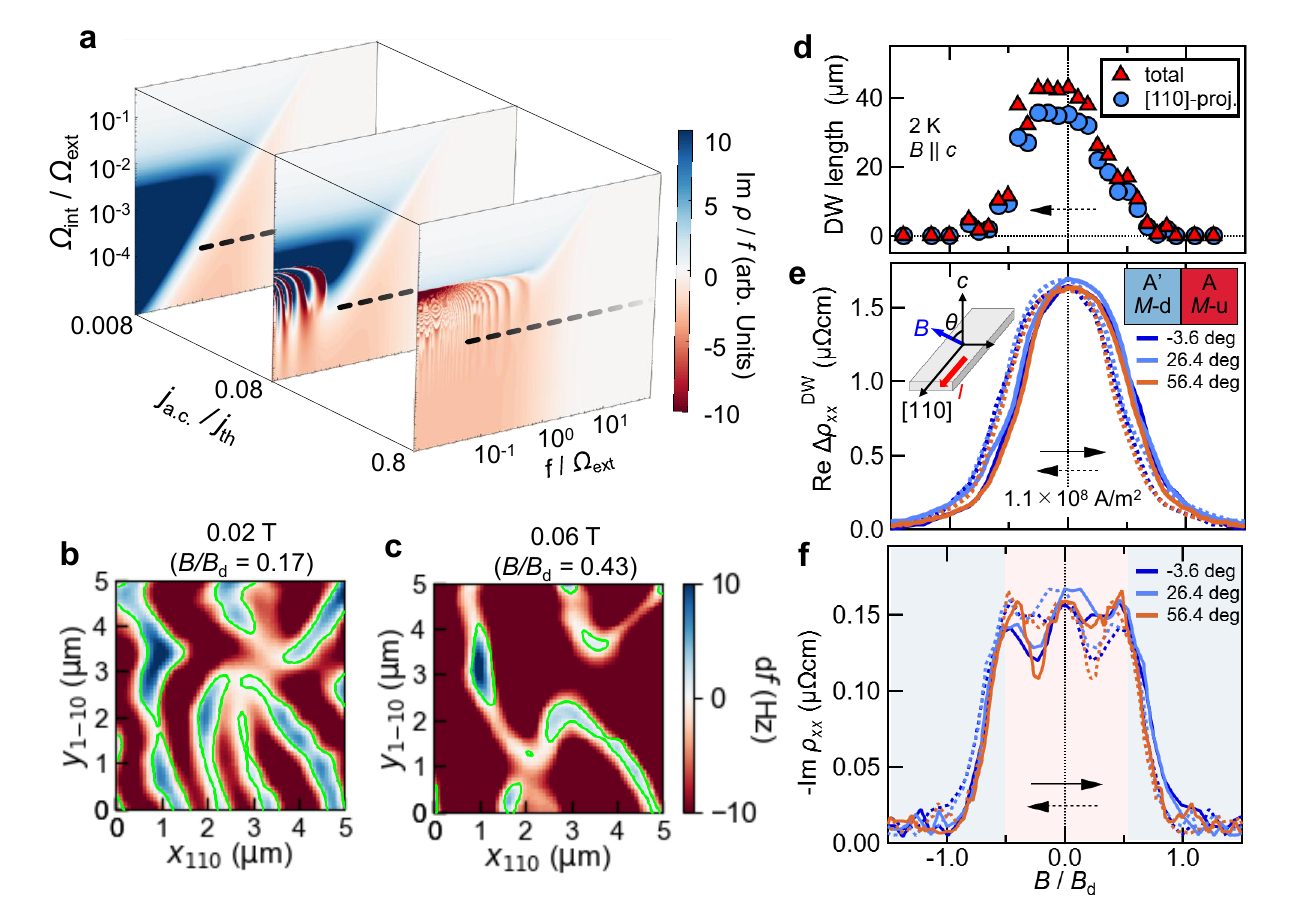}
\caption{\textbf{Correlation of domain wall (DW) density and emergent electric field (EEF) from DW motion.} $\mathbf{a,}$ Imaginary part of the complex impedance $\Im\rho_{xx}$ divided by frequency $f$ as obtained from numerical calculations of the EEF in our spin dynamics model for a single magnetic DW. $\Omega_\mathrm{int}$ and $\Omega_\mathrm{ext}$ are defined in Eq.~(\ref{EqofMotion_X_twodots}), while $j_\mathrm{th}$ corresponds to the depinning threshold of the tilting mode in the $\Omega_\mathrm{int} \ll \Omega_\mathrm{ext}$ limit. A negative imaginary impedance, or negative EEF, appears over a wide range of $f$ when $\Omega_\mathrm{int} / \Omega_\mathrm{ext}$ is small, i.e., where spin-tilting $\phi$ has a low energy cost. This corresponds to DW motion dominated by friction, as in Fig.~\ref{Fig1}\textbf{e}, where the potential term in Eq.~(\ref{EqofMotion_X_twodots}) becomes irrelevant to the dynamics. In the regime below the black dotted line, the model predicts negative $\Im\rho_{xx}$ over a wide parameter range. $\mathbf{b,c,}$ MFM images taken during a field scan from positive to negative field (field-down sweep). When the magnetic field is close to $B_\mathrm{d}$, the size of the domain with negative net magnetization is enhanced, and accordingly the number of DWs decreases. $\mathbf{d,}$ Magnetic field dependence of the DW length. The red triangles (blue circles) show the total DW length (projection of the DW length onto the $\langle 110 \rangle$ axis) corresponding to the green lines in panels $\mathbf{b,c}$, respectively. $\mathbf{e,f,}$ The DW-resistivity $\Re\rho_{xx}^\mathrm{DW}$ and the imaginary impedance $\Im\rho_{xx}$ when the magnetic field is tilted away from the vertical axis by an angle $\theta$. The horizontal axis is the magnetic field normalised by $B_\mathrm{d}$. \textcolor{black}{The blue and red shading in panel \textbf{f} indicates the regimes where changes in DW population and changes in DW pinning are dominant, respectively (see also Fig.~4).} 
}\label{Fig3}
\end{figure}

We observe an imaginary impedance of negative sign over a wide range of frequency and current density (Extended Data Fig. 4). 
To explain the negative sign of the observed imaginary impedance, we numerically calculate the EEF and $\Im \rho_{xx}$ for a single DW between two magnetic domains. Figure \ref{Fig3}\textbf{a} visualises distinct regimes of the model in terms of the dimensionless ratios $\Omega_\mathrm{int} / \Omega_\mathrm{ext}$ and $f / \Omega_\mathrm{ext}$. The overall behaviour depends on the current density applied to the device, but we see that a $\Im\rho_{xx}$ of positive sign appears when $\Omega_\mathrm{int}$ approaches the order of $\Omega_\mathrm{ext}$, while $\Im\rho_{xx}$ takes negative values for a wide range of $f / \Omega_\mathrm{ext}$ when $\Omega_\mathrm{int} / \Omega_\mathrm{ext} \ll 1$. Since $\Omega_\mathrm{int}$ is the energy cost associated with spin tilting $\phi$,
$\Omega_\mathrm{int} / \Omega_\mathrm{ext} \ll 1$ means that the energy scale of DW sliding $X$ dominates the dynamical properties of the current-driven DW; see
Extended Data Fig. 5 and Supplementary Note 6. 

The experimental observation of a negative $\Im\rho_{xx}$ of DWs over a wide range of parameters, in combination with our numerical calculations, thus heralds an important role of friction in DW dynamics of NdAlSi: According to Eq.~(\ref{EqofMotion_X_twodots}), the effective pinning potential of the sliding mode grows feebler and feebler \textcolor{black}{with decreasing $\Omega_\mathrm{int}$} so that small-amplitude motion of $X$ is realized with only driving force, kinetic energy, and friction terms. 
The dissipative nature of this response is demonstrated by Cole-Cole analysis in Supplementary Fig. 2 and Supplementary Note 5, without relying on details of a theoretical model. \\

\textbf{Dynamics and EEF of domain walls in magnetic field}\\
We analyse the dual role of the external magnetic field $B$ on the emergent electric field (EEF), focusing on two key aspects: (i) the variation in domain wall (DW) length, and (ii) the modulation of the DW pinning potential. The DW length decreases when the magnetic field takes large positive or negative values, as shown in Figs.~\ref{Fig2}\textbf{b} and \ref{Fig3}\textbf{b,c} (see also Extended Data Fig. 6). Figure \ref{Fig3}$\mathbf{d}$ shows the total DW length, and the DW length projected onto the normal to the current direction, as extracted from MFM images with a field of vision of $5 \times 5 \,\mathrm{\mu m}$ (see Supplementary Note 1). When decreasing the magnetic field starting from large positive values, both total and projected DW lengths increase below $B_\mathrm{d}$; they take a maximum around $B=0$, with a small hysteresis. In Fig.~\ref{Fig3}\textbf{e,f}, this change of DW length is reflected in both real and imaginary parts of the complex impedance, $\Re \rho_{xx}$ and $\Im\rho_{xx}$, indicating that both are enhanced when larger DW length is realized within a device. Tilting the magnetic field by an angle $\theta$ away from the normal axis of the device, we confirm that $B_\mathrm{d}$ follows a $1/\cos\theta$ dependence, as expected for the threshold field of domain formation in magnets with predominantly Ising character (see Extended Data Fig. 7 and Supplementary Note 4). Corrected for this change of $B_\mathrm{d}$, $\Re \rho_{xx}^\mathrm{DW}$ -- with its bell shape and \textcolor{black}{its} maximum around zero field -- is regarded as a good measure of DW population. 

\begin{figure}[h]%
\centering
\includegraphics[trim=0cm 0.1cm 0.0cm 0.0cm,clip,width=0.7\textwidth]{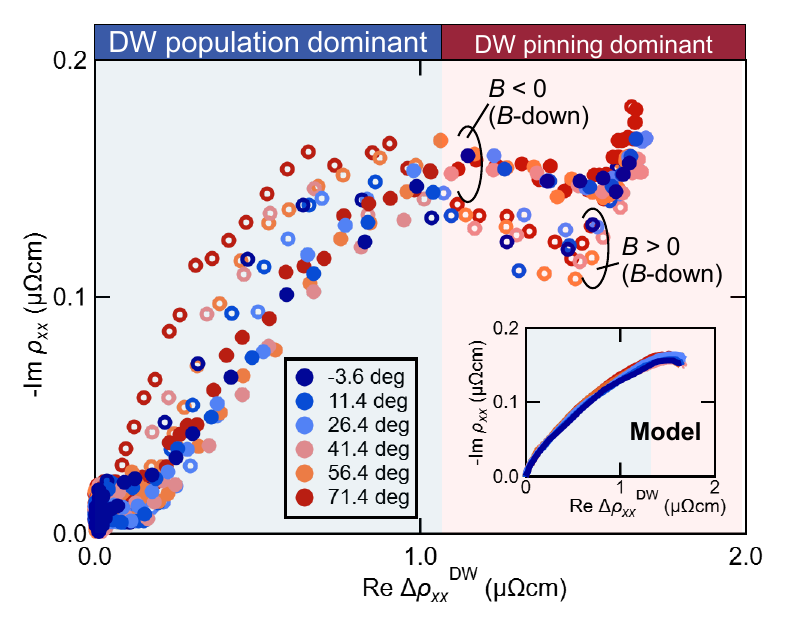}
\caption{\textbf{Scaling of domain wall (DW) resistivity and the imaginary impedance.} $\Re\rho_{xx}^\mathrm{DW}$ and $\Im\rho_{xx}$, as defined in Fig.~\ref{Fig3}, from field down-sweeps for $B>0$ and $B<0$ are shown as open and filled circles, respectively. Close to $B_\mathrm{d}$, the two quantities show near-linear correlation \textcolor{black}{(``DW population dominant'' regime, blue shading)}; but $\Im\rho_{xx}$ exhibits a plateau when $\Re \rho_{xx}^\mathrm{DW}$ is still rising, due to a $B$-dependent change in the EEF beyond the domain wall density \textcolor{black}{(``DW pinning dominant'' regime, red shading)}. The inset shows our EEF model calculation for $\Im\rho_{xx}$ with Gilbert damping $\alpha=10^{\mathchar`-3}\sim10^{\mathchar`-4}$ (see text for discussion).
}\label{Fig4}
\end{figure}

Within the blue area around $B_\mathrm{d}$ in Fig.~\textcolor{black}{\ref{Fig3}\textbf{f} and} \ref{Fig4}, $\Im \rho_{xx}$ increases in proportion to $\Re \rho_{xx}^\mathrm{DW}$. We expect both quantities to be dominated by dramatic changes in DW population in this field range. At lower $|B|$, or larger $\Re \rho_{xx}^\mathrm{DW}$, $\Im \rho_{xx}$ saturates. This observed saturation of $\Im \rho_{xx}$ in the red region of Fig.~\ref{Fig3}\textbf{f} \textcolor{black}{and \ref{Fig4}} follows readily from our theoretical model, where the EEF --- but not $\Re \rho_{xx}^\mathrm{DW}$ nor the DW length --- is enhanced by a suppression of the pinning frequency $\Omega_\mathrm{ext} $ with the square of the external field, $B^2$. More specifically, the magnetic field induces changes in the volume fraction of two magnetic domains and thus shifts the DW position away from its original pinning centre. This results in a suppression of effective pinning strength, larger amplitude of oscillations in $X$, and larger EEF. In fact, the decrease in DW population and enhanced EEF from the change in pinning strength nearly compensate, producing flat $\Im \rho_{xx}(B)$ traces at low magnetic fields. In Fig.~\ref{Fig4}, inset, we describe the $\Im\rho_{xx}$ vs. $\Re\rho_{xx}^\mathrm{DW}$ trend in Fig.~\ref{Fig4} using our numerical model (Methods). We extract the Gilbert damping parameter $\alpha=10^{\mathchar`-3}\sim10^{\mathchar`-4}$, a reasonable value for semimetallic magnets~\cite{JSeib2010, SMankovsky2013}.
We note that a dip in $\Im\rho_{xx}$ at $B > 0$ originates from different domain patterns as discussed in Extended Data Fig. 8 and Supplementary Note 2.\\

\textbf{Discussion}\\
The dissipative dynamics of pinned DWs under an a.c.~drive should be observed widely when the excitation frequency is high enough and Dzyaloshinskii-Moriya or dipolar interactions are sufficiently weak, as compared to the saturation magnetization [after converting to suitable units, see Eq.~(\ref{EqofIntrinsic_freq})]. Realized in NdAlSi, this limit also entails an effective depinning of the tilt mode $\phi$, Fig.~\ref{Fig1}\textbf{c}, which becomes directly dependent on $X$ and acquires a large oscillation amplitude under current drive [Eq.~(\ref{LinEqMotion_phi}) of Methods]. According to Nagaosa~\cite{NNagaosa2019}, the figure of merit $Q$ is expected to be proportional to the excitation frequency $f$ and the electric conductivity $\sigma$, while it is inversely proportional to $\Omega_\mathrm{int}$ and the domain wall width $\lambda$. This provides a rationale for the large $\Im\rho_{xx}$ in NdAlSi, despite a small volume fraction of DWs. \\

\color{black}
\textbf{Conclusion and Outlook}\\
Looking ahead, our results on the EEF response and energy dissipation in domain wall dynamics open opportunities to investigate diverse energy dissipation mechanisms in domain walls~\cite{Bouzidi1990, Stamp1991, Braun1996, Maho2009, Brataas2011, Kim2018}, and to extend the study of EEFs to solitons with more complex dynamics, such as vortices and skyrmions. In particular, we expect an impact of a vortex's gyrating (circular) motion on the EEF~\cite{NNagaosa2012}. Furthermore, the present approach provides a foundation for probing emergent electrodynamics under extreme conditions, including regimes near quantum criticality or where domain-wall quantum fluctuations become significant~\cite{Chudnovsky1992,Braun1997,Andreas2022}.
\\
\color{black}

\textbf{Acknowledgements}\\
We acknowledge M. Birch, I. Belopolski, P. R. Baral,
G. Chang, S. Sen, A. Ozawa, 
N. Nagaosa and T.-h. Arima for fruitful discussion, and RIKEN CEMS Semiconductor Science Research Support Team for technical assistance. 
\color{black}
R. Yamada was supported by JSPS KAKENHI Grants No. 22K20348, No. 23K13057, No. 24H01604, and No. 25K17336, and JST PRESTO JPMJPR259A, as well as the Foundation for Promotion of Material Science and Technology of Japan, the Yashima Environment Technology Foundation, the Yazaki Memorial Foundation for Science and Technology, and the ENEOS Tonengeneral Research/Academic Foundation.
Y. Fujishiro was supported by JST PRESTO JPMJPR2597.
S. Okumura was supported by JSPS KAKENHI Grants No. 22K13998 and No. 23K25816, and JST PRESTO JPMJPR2595.
T. Yokouchi was supported by JSPS KAKENHI Grants No. 24K00566 and JST PRESTO JPMJPR235B.
Y. Tokura was supported by JSPS KAKENHI Grants No. 23H05431 and JST CREST Grant Number JPMJCR1874.
M. Hirschberger was supported by JSPS KAKENHI Grants No. 21K13877, No. 22H04463, No. 23H05431, and No. 24H01607, and JST CREST Grant Number JPMJCR20T1 and JST FOREST Grant No. JPMJFR2238, as well as the Fujimori Science and Technology Foundation, the New Materials and Information Foundation, the Murata Science Foundation, the Mizuho Foundation for the Promotion of Sciences, the Yamada Science Foundation, the Hattori Hokokai Foundation, the Iketani Science and Technology Foundation, the Mazda Foundation, the Casio Science Promotion Foundation, the Takayanagi Foundation, and Inamori Foundation. M. H. is also supported by Japan Science and Technology Agency (JST) as part of Adopting Sustainable Partnerships for Innovative Research Ecosystem (ASPIRE), Grant Number JPMJAP2426. M.H. is supported by the Deutsche Forschungsgemeinschaft (DFG, German Research Foundation) via Transregio TRR 360 – 492547816.
O.A. Tretiakov acknowledges the support from the Australian Research Council (Grant Nos. DP200101027 and DP240101062) and the NCMAS grant. Research sponsored by the Laboratory Directed Research and Development Program of Oak Ridge National Laboratory, managed by UT-Battelle, LLC, for the US Department of Energy.\\
\color{black}

\textbf{Author Contributions}\\
YTa, YTo, and MH conceived the project. RY and AK synthesised the bulk NdAlSi single crystal. RY and YF fabricated the focused ion beam devices. RY measured MFM images and analysed data with support from DN and FSY. RY performed the transport measurement and analysis with support from DK, TY, and MH. DK and OAT developed the theoretical model and performed numerical calculations. RY and TN performed polarised neutron scattering measurements. SO performed spin model calculations. RY and MH wrote the manuscript with the contributions from all other coauthors.\\

\textbf{Competing interests}\\
The authors declare no competing interests.\\


\newpage

\clearpage
\begin{center}
\Large{Main text Figures}
\end{center}

\clearpage

\clearpage

\clearpage

\makeatletter
\newcounter{prevbib}
\setcounter{prevbib}{\value{NAT@ctr}}
\makeatother

\newpage

\begin{center}
\Large{Methods}
\end{center}

\textbf{Device fabrication}\\
Single crystalline samples of NdAlSi were synthesised by the self-flux method~\cite{JGaudet2021, RYamada2024}. The orientation of the crystals was confirmed by Laue X-ray diffraction patterns. We cut oriented, micrometer-sized devices of NdAlSi using the FIB technique (NB-5000, Hitachi). The devices were mounted on a silicon substrate and fixed by FIB-assisted tungsten deposition. A scanning electron microscope (SEM) image of a mesoscopic device is shown in Fig.~\ref{Fig2}\textbf{c}. Electrical contact to the sample is made by FIB-assisted tungsten (W) deposition. We carefully removed any residual tungsten on the surface of the device to avoid an\textcolor{black}{y} effect of the superconductivity of amorphous tungsten, resulting in black grooves visible in Fig.~\ref{Fig2}\textbf{c}. 
The surface of the device used in the Main Text is perpendicular to the crystallographic $c$ axis of tetragonal NdAlSi, 
and the magnetic field $B$ (the current $I_\mathrm{a.c.}$) is applied parallel to the $c$ axis (along the $[110]$ axis). \textcolor{black}{The sample temperature with a finite current density is estimated from the value of the magnetic field where the ferrimagnetic state transitions to the ferromagnetic state (this critical field has a sizable temperature dependence).}\\

\textbf{Transport measurements}\\
Measurements of the complex resistivity are performed by the lock-in technique (LI5650, NF Corporation) using a standard four-probe method. Using an FIB-fabricated device of thickness $t\sim0.8\,\mathrm{\mu m}$, we apply a sine-wave current and detect both in-phase (Re $V_\mathrm{a.c.}^{1f}$) and out-of-phase (Im $V_\mathrm{a.c.}^{1f}$) components of the voltage. The extrinsic rotation of the phase in the lock-in measurements, due to stray capacitance of the circuit, was subtracted by assuming that Im $V_\mathrm{a.c.}^{1f}$ is zero in the collinear, field-aligned state. The extrinsic phase rotation depends on frequency, but it is below $0.1\,$degrees at and around $f_0 \sim 17 \, \mathrm{Hz}$, and has a weak temperature dependence.\\

\textbf{Magnetic force microscopy (MFM)}\\
Magnetic force microscopy (MFM) measurements were conducted with a scanning probe microscope (attocube AFM/MFM I). The detection mechanism is based on phase interference of an optical beam injected into the backside of the cantilever, which is oscillated by applying an a.c.~voltage to a piezoelectric dither motor. The edge of the optical fiber is located a few millimeters above the backside of the cantilever. The shift of the resonance frequency of the cantilever, due to the change in the stray magnetic field from the sample surface, is detected by a shift in the interference pattern between the reflected light and a reference beam.

We recorded MFM data for both bulk and device samples, where the plane of the crystal is perpendicular to the crystallographic $c$ axis. The DW patterns are reproducible without a significant impact of dipolar energy for bulk samples (thickness $60\sim300\,\mathrm{\mu m}$) and FIB-fabricated devices ($3\sim6\,\mathrm{\mu m}$). We show MFM images of FIB-fabricated devices with a thickness of $6\,\mathrm{\mu m}$ in Figs.~\ref{Fig2}\textbf{b} and \ref{Fig3}\textbf{b,c} measured after field-cooling the device in a magnetic field of $6\,\mathrm{T}$ and then decreasing the magnetic field to zero. In order to compare the field dependence between devices of different dimensions, we performed a demagnetization correction, where the sample was assumed to have the shape of a flat ellipsoid~\cite{AAharoni1998}.\\

\color{black}

\textbf{Modeling of emergent electric field (EEF) from domain walls (DWs)}\\
We consider the following magnetic free energy model
\begin{equation}
\label{EqofMagnetic_free_energy}
F[\bm{m}] = S \int dx\left[ A \left( \frac{\partial \bm{m}}{\partial x} \right)^2 + D \left( \bm{m} \times \frac{\partial \bm{m}}{\partial x} \right)_x - K m_z^2 + a K_\mathrm{p} \delta(x)m_z^2 \right],
\end{equation}
where $S$ is the sample cross-section and $a$ is the lattice constant. The first, second, third, and fourth terms correspond to the exchange energy, Dzyaloshinskii-Moriya (DM) interaction forming N{\'e}el-type DWs, easy-axis anisotropy, and extrinsic (defect) pinning potential, respectively. Here we note that the DM interaction is allowed in the polar crystal structure of NdAlSi for nearest neighbor Nd sites along the $\langle 110 \rangle$ direction. The main results are unchanged when using Bloch-type DMI, because the present EEF model is based on spin-transfer torque. This magnetic free energy has a DW solution described by
\begin{equation}
\label{EqofDW_solution}
\bm{m}(x, t) = \left[ \frac{\cos{\phi}(t)/2}{\cosh{\frac{x-X(t)}{2 \lambda}}},  \frac{\sin{\phi}(t)/2}{\cosh{\frac{x-X(t)}{2 \lambda}}}, \tanh{\frac{x-X(t)}{2 \lambda}}\right],
\end{equation}
where $\lambda = \frac{1}{2} \sqrt{A/K}$ is the equilibrium domain wall width.

The equations of motion under the excitation current for the two canonically conjugated, collective variables -- position $X$ and spin tilting angle $\phi$ -- are 
\begin{equation}
\label{EqofMotion_X}
\dot{X}+2\Omega_\mathrm{int}\sin{\left(\frac{\phi}{2}\right)} + \frac{2\alpha \Omega_\mathrm{ext} \tanh{\left(\frac{X}{2}\right)}}{\cosh^2{\left(\frac{X}{2}\right)}} - \frac{Pg\mu_\mathrm{B}}{2e\lambda M_s} \frac{1 + \alpha \beta}{1 + \alpha^2}j_\mathrm{a.c.} = 0
\end{equation}
\begin{equation}
\label{EqofMotion_phi}
\dot{\phi}+2\alpha\Omega_\mathrm{int}\sin{\left(\frac{\phi}{2}\right)} - \frac{2\Omega_\mathrm{ext} \tanh{\left(\frac{X}{2}\right)}}{\cosh^2{\left(\frac{X}{2}\right)}} - \frac{Pg\mu_\mathrm{B}}{2e\lambda M_s} \frac{\alpha - \beta}{1 + \alpha^2}j_\mathrm{a.c.} = 0
\end{equation}
Here, $\Omega_\mathrm{int}$, $\Omega_\mathrm{ext}$, $\alpha$, $\beta$, $P$, $g$, $\mu_\mathrm{B}$, $e$, and $M_s$ are the intrinsic pinning frequency, extrinsic pinning frequency, Gilbert damping constant, a dimensionless constant characterising the nonadiabatic spin-transfer torque, the spin polarization of conduction electrons, electronic $g$-factor, Bohr magneton, elementary charge, and saturation magnetization, respectively. The current density $j_\mathrm{a.c.}$ is defined as the electric current divided by the sample's cross-sectional area, $j_\mathrm{a.c.} = I_\mathrm{a.c.}/S$. The intrinsic ($\Omega_\mathrm{int}$) and extrinsic ($\Omega_\mathrm{ext}$) pinning frequencies are given by
\begin{equation}
\label{EqofIntrinsic_freq}
\Omega_\mathrm{int} = \frac{\pi \gamma D}{4(1+\alpha^2)\lambda \mu M_\mathrm{s}},
\end{equation}
\begin{equation}
\label{EqofExtrinsic_freq}
\Omega_\mathrm{ext} = \frac{a \gamma K_\mathrm{p}}{2(1+\alpha^2)\lambda \mu M_\mathrm{s}}.
\end{equation}
where $\gamma$ and $\mu$ are the gyromagnetic ratio and the permeability of vacuum.

Next, we calculate the EEF from the spin-motive force, which is described by~\cite{NNagaosa2019}
\begin{equation}
\label{EqofSMF}
V(t) = - \frac{\hbar P}{2 e} \int dx \left( \bm{n} \times \frac{\partial \bm{n}}{\partial t} + \beta \frac{\partial \bm{n}}{\partial t} \right) \cdot \frac{\partial \bm{n}}{\partial x} = - \frac{\hbar P}{2 e} \left( \frac{d \phi(t)}{dt} - \beta \frac{d X(t)}{dt}\right).
\end{equation}
Finally, the first harmonic response of the imaginary impedance $\mathrm{Im} \, \rho_{xx} = \mathrm{Im} \, V^{1f}/lj_\mathrm{a.c.}$ in the limit of $\Omega_\mathrm{int} \ll \Omega_\mathrm{ext}$, is obtained as 
\begin{align}
\label{EqofImImpedance_full}
\frac{\mathrm{Im} \, \rho_{xx}l}{Sf} = -\left( \frac{\hbar P}{4e S} \right) \Big[& \frac{g \mu_\mathrm{B} P}{e M_s \lambda} \frac{(1 + \alpha \beta)^2}{\alpha(1 + \alpha^2)} \frac{\Omega_\mathrm{ext}}{(f^2 + \Omega_\mathrm{ext}^2)} +\notag\\
&\left( \frac{g \mu_\mathrm{B} P}{e M_s \lambda} \right)^3 \frac{(1 + \alpha \beta)^4}{4 \alpha (1 + \alpha^2)^3} \frac{(f^2 - \Omega_\mathrm{ext}^2)}{(f^2 + \Omega_\mathrm{ext}^2)^3} j_\mathrm{a.c.}^2 \Big].
\end{align}
Here, the first and second terms represent the linear ($j_\mathrm{a.c.}^0$) and leading non-linear ($j_\mathrm{a.c.}^2$) contributions, respectively. 

To simplify the discussion, the equations of motion, Eqs.~(\ref{EqofMotion_X},\ref{EqofMotion_phi}), can be linearised in terms of $X$ and $\phi$, when the current drive is weak enough and the distortions $\delta X$ and $\delta \phi$ are small enough:
\begin{equation}
\label{LinEqMotion_X}
\dot{X} + \Omega_\mathrm{int}\phi + \alpha\Omega_\mathrm{ext} X - \frac{1 + \alpha \beta}{1 + \alpha^2} \nu = 0
\end{equation}
\begin{equation}
\label{LinEqMotion_phi}
\dot{\phi} + \alpha\Omega_\mathrm{int} \phi - \Omega_\mathrm{ext} X - \frac{\alpha - \beta}{1 + \alpha^2} \nu = 0
\end{equation}
Here, $\nu=j_\mathrm{a.c.} Pg\mu_\mathrm{B}/2e\lambda M_s$ is proportional to $j_\mathrm{a.c.}$. \textcolor{black}{We discuss the applicability of our spin model to NdAlSi in Supplementary Table 2, Supplementary Note 7, and Supplementary Note 8.} \textcolor{black}{We discuss possible contributions from spin-orbit torque by using a device with a different configuration in Extended Data Fig. 3 and Supplementary Note 10~\cite{JIeda2021, YYamane2022}. Finally, we also note that enhanced spin transfer torque by a charged DW is proposed for magnetic Weyl semimetals~\cite{DKurebayashi2019}, however, such contributions may be screened by conduction electrons in NdAlSi. }\\

\textbf{Model for emergent electric field under a magnetic field}\\
Here we consider the effect of external magnetic fields $B$ on the emergent eclectic field (EEF) from magnetic DWs. When $B$ is along the easy axis of magnetization ($z$ axis), the equations of motion, Eqs.~(\ref{EqofMotion_X},\ref{EqofMotion_phi}), are modified as follows:
\begin{equation}
\label{EqofMotion_X_Bfield}
\dot{X}+2\Omega_\mathrm{int}\sin{\left(\frac{\phi}{2}\right)} + \frac{2\alpha \Omega_\mathrm{ext} \tanh{\left(\frac{X}{2}\right)}}{\cosh^2{\left(\frac{X}{2}\right)}} - \frac{Pg\mu_\mathrm{B}}{2e\lambda M_s} \frac{1 + \alpha \beta}{1 + \alpha^2}j_\mathrm{a.c.} + \frac{2 \alpha \gamma B}{1 + \alpha^2} = 0
\end{equation}
\begin{equation}
\label{EqofMotion_phi_Bfield}
\dot{\phi}+2\alpha\Omega_\mathrm{int}\sin{\left(\frac{\phi}{2}\right)} - \frac{2\Omega_\mathrm{ext} \tanh{\left(\frac{X}{2}\right)}}{\cosh^2{\left(\frac{X}{2}\right)}} - \frac{Pg\mu_\mathrm{B}}{2e\lambda M_s} \frac{\alpha - \beta}{1 + \alpha^2}j_\mathrm{a.c.} - \frac{2 \alpha \gamma B}{1 + \alpha^2}= 0
\end{equation}
For the equilibrium solution ($j_\mathrm{a.c.} = 0$), we obtain $\phi_0 = 0$ and $X_0 \sim -2\gamma B / \Omega_\mathrm{ext} + \mathcal{O}(B^3)$. This result corresponds to the fact that the centre of the effective pinning potential shifts under the magnetic field due to the change in the volume fraction of the two magnetic domains with different net magnetization. By linearising Eqs.~(\ref{EqofMotion_X_Bfield},\ref{EqofMotion_phi_Bfield}) around the equilibrium solution in terms of $X$ and $\phi$, we derive
\begin{equation}
\label{EqofMotion_X_BfieldLin}
\dot{X}+\Omega_\mathrm{int}\phi + \alpha\Omega_\mathrm{ext}^\mathrm{mod}X - \frac{Pg\mu_\mathrm{B}}{2e\lambda M_s} \frac{1 + \alpha \beta}{1 + \alpha^2}j_\mathrm{a.c.} = 0
\end{equation}
\begin{equation}
\label{EqofMotion_phi_BfieldLin}
\dot{\phi}+\alpha\Omega_\mathrm{int}\phi-\Omega_\mathrm{ext}^\mathrm{mod}X - \frac{Pg\mu_\mathrm{B}}{2e\lambda M_s} \frac{\alpha - \beta}{1 + \alpha^2}j_\mathrm{a.c.}= 0
\end{equation}
Here, $\Omega_\mathrm{ext}^\mathrm{mod} = \Omega_\mathrm{ext}\left[1 - \frac{4 \gamma^2 B^2}{(1 + \alpha^2)^2 \Omega_\mathrm{ext}^2} \right]$. Finally, using these modified equations of motion under the magnetic field, we can derive the imaginary impedance for magnetic field along the $z$ axis in the limit of small $\Omega_\mathrm{int}$ and $f$ as 
\begin{equation}
\frac{\mathrm{Im} \, \rho_{xx}l}{Sf} = \left(\frac{\mathrm{Im} \, \rho_{xx}l}{Sf}\right)_1 + \left(\frac{\mathrm{Im} \, \rho_{xx}l}{Sf}\right)_2
\end{equation}
where
\begin{equation}
\label{ImImp_Bfield_lin}
\left(\frac{\mathrm{Im} \, \rho_{xx}l}{Sf}\right)_1 = -\left( \frac{\hbar P}{4e S} \right) \frac{g \mu_\mathrm{B} P}{e M_s \lambda} \frac{(1 + \alpha \beta)^2}{\alpha^2(1 + \alpha^2)} \frac{1}{\Omega_\mathrm{ext}}\left(1 + 4\gamma^2 \frac{1}{(1 + \alpha^2)^2} \frac{B^2}{\Omega_\mathrm{ext}^2} \right)
\end{equation}
\begin{equation}
\label{ImImp_Bfield_nonlin}
\left(\frac{\mathrm{Im} \, \rho_{xx}l}{Sf}\right)_2 = -\left( \frac{\hbar P}{16e S} \right) \left( \frac{g \mu_\mathrm{B} P}{e M_s \lambda} \right)^3 \frac{(1 + \alpha \beta)^4}{\alpha^4 (1 + \alpha^2)^3}  \frac{1}{\Omega_\mathrm{ext}^3} \left( 1 + \frac{63}{2} \gamma^2 \frac{1}{(1 + \alpha^2)^2}\frac{B^2}{\Omega_\mathrm{ext}^2}\right) j_\mathrm{a.c.}^2
\end{equation}
Therefore, the negative imaginary impedance is enhanced under the magnetic field due to a reduction of effective pinning frequency, as long as the DWs are in the pinned regime. When the magnetic field is strong enough to depin the DWs from the local defect, and to ultimately destroy the minority domains, the number of DWs and the imaginary impedance $\Im\rho_{xx}$ decrease with magnetic field. This is experimentally observed in Fig.~\ref{Fig3}\textbf{f}.\\

\textbf{Scaling relation between DW resistivity and imaginary impedance}\\
As demonstrated in Fig.~\ref{Fig3}\textbf{d,e}, the field dependence of the DW resistivity $\mathrm{Re} \, \Delta \rho_{xx}^\mathrm{DW}$ is closely correlated with the density of the domain walls $N^\mathrm{DW}$. Therefore, we estimate the DW length via the DW resistivity $\mathrm{Re} \, \Delta \rho_{xx}^\mathrm{DW}$ using a constant of proportionality $C_{1}$ as
\begin{equation}
\label{EqofDWRes_DWdensity}
\mathrm{Re} \, \Delta \rho_{xx}^\mathrm{DW} = C_{1} N^\mathrm{DW}.
\end{equation}

The imaginary impedance is induced by the magnetic DWs, therefore it is also expected to scale with $N^\mathrm{DW}$. However, an additional contribution from the magnetic field appears in $\mathrm{Im} \, \rho_{xx}$ as shown in Eqs.~(\ref{ImImp_Bfield_lin},\ref{ImImp_Bfield_nonlin}). Phenomenologically, the field dependence of the imaginary impedance is expressed as
\begin{equation}
\label{EqofImImpedance_DWdensity}
\mathrm{Im} \, \rho_{xx} = C_{2} N^\mathrm{DW} (1 + C_{3}B^2),
\end{equation}
where $C_{2}$ and $C_{3}$ are constants. We note that the DW resistivity is governed by the scattering of conduction electrons by the magnetic DWs according to Eq.~(\ref{EqofDWRes_DWdensity}), and the magnetic field effect on $\mathrm{Re} \, \Delta \rho_{xx}^\mathrm{DW}$ is higher-order than that on $\mathrm{Im} \, \rho_{xx}$. Therefore, the scaling relation between DW resistivity and imaginary impedance is given as follows:
\begin{equation}
\label{EqofDWRes_ImImpedance}
\mathrm{Im} \, \rho_{xx} = k \mathrm{Re} \, \Delta \rho_{xx}^\mathrm{DW} (1  + l B^2),
\end{equation}
where $k$ and $l$ are constants independent of the magnetic field. We fit the experimental data to the model in Eq.~(\ref{EqofDWRes_ImImpedance}) and show the fitting results in the inset to Fig.~\ref{Fig4}. The fit reproduces the nearly linear relation between $\mathrm{Im} \, \rho_{xx}$ and $\mathrm{Re} \, \Delta \rho_{xx}^\mathrm{DW}$ when the DW population is small, and also the plateau-like feature in the regime of large DW population. From the fitting, we extract to be $k = -0.092$ and $l = C_3 = 89 \, \mathrm{T}^{-2}$, respectively. Considering Eqs.~(\ref{ImImp_Bfield_lin},\ref{ImImp_Bfield_nonlin}), we reproduce the experimentally extracted parameter $C_3$ by setting the Gilbert damping constant to $\alpha=10^{\mathchar`-3}\sim10^{\mathchar`-4}$. \\

\end{document}